\pdfoutput=1
\documentclass{nature}  
\usepackage{relsize}

\usepackage{amsmath,amsfonts,amssymb}
\usepackage{graphicx}

\usepackage{enumitem} 
\setlist{partopsep=-\parskip,parsep=0pt}

\usepackage{sistyle}\SIdecimalsign{.}
\usepackage[Euler]{upgreek}
\newcommand*{\micro}{\ensureupmath{\upmu}}

\newcommand{\dd}[1][]{\text{d{#1}}} 
\newcommand{\F}{\mathcal{F}} 

\usepackage[capitalise]{cleveref}
\usepackage{xcolor}

\usepackage{epstopdf}

\usepackage{lineno}
\newcommand*\patchAmsMathEnvironmentForLineno[1]{
	\expandafter\let\csname old#1\expandafter\endcsname\csname #1\endcsname
	\expandafter\let\csname oldend#1\expandafter\endcsname\csname end#1\endcsname
	\renewenvironment{#1}
	{\linenomath\csname old#1\endcsname}
	{\csname oldend#1\endcsname\endlinenomath}}
\newcommand*\patchBothAmsMathEnvironmentsForLineno[1]{
	\patchAmsMathEnvironmentForLineno{#1}
	\patchAmsMathEnvironmentForLineno{#1*}}
\AtBeginDocument{
	\patchBothAmsMathEnvironmentsForLineno{equation}
	\patchBothAmsMathEnvironmentsForLineno{align}
	\patchBothAmsMathEnvironmentsForLineno{flalign}
	\patchBothAmsMathEnvironmentsForLineno{alignat}
	\patchBothAmsMathEnvironmentsForLineno{gather}
	\patchBothAmsMathEnvironmentsForLineno{multline}
}

\usepackage[font={footnotesize}]{caption}

\title{Diffraction-limited axial scanning in thick biological tissue employing an aberration correcting adaptive lens}

\author{Katrin Philipp$^{1,*}$, Florian Lemke$^2$, Stefan Scholz$^3$, Ulrike Wallrabe$^2$, Matthias C. Wapler$^2$, Nektarios Koukourakis$^1$, J\"urgen W. Czarske$^1$}

\usepackage{graphicx}
\makeatletter
\let\saved@includegraphics\includegraphics
\AtBeginDocument{\let\includegraphics\saved@includegraphics}
\renewenvironment*{figure}{\@float{figure}}{\end@float}
\makeatother

\begin{document} 
\relsize{-1}

\maketitle

\begin{affiliations}
	\item Technische Universit{\"a}t Dresden, Laboratory for Measurement and Sensor System Techniques,	Helmholtzstra{\ss}e 18, 01069 Dresden, Germany
	\item University of Freiburg, Laboratory for Microactuators, Department of Microsystems Engineering-IMTEK, Georges-K\"ohler-Allee 102, 79110 Freiburg, Germany
	\item Helmholtz Centre for Environmental Research–UFZ, Department of Bioanalytical Ecotoxicology, Leipzig, Germany\\
	* katrin.philipp@tu-dresden.de
\end{affiliations}

\begin{abstract}
Diffraction-limited focusing deep into biological tissue is challenging due to spherical aberrations that lead to a broadening of the focal spot particularly in axial direction. While the diffraction-limit can be restored employing aberration correction with a deformable mirror or spatial light modulator, a bulky optical setup results due to the required beam-folding. We propose a bi actor adaptive lens, that enables axial scanning and at the same time correction of specimen induced spherical aberrations while offering a compact setup.  Using the bi-actor lens in a confocal microscope, we show diffraction limited axial scanning up to \SI{340}{\micro m} deep inside a phantom specimen. Applying this technique for in-vivo measurements of zebrafish embryos with reporter gene-driven fluorescence in the thyroid gland reveals substructures of thyroid follicles.
\end{abstract}

\section{INTRODUCTION}
\label{sec:intro}  

\begin{figure}
	\centering
	\includegraphics[width=\linewidth]{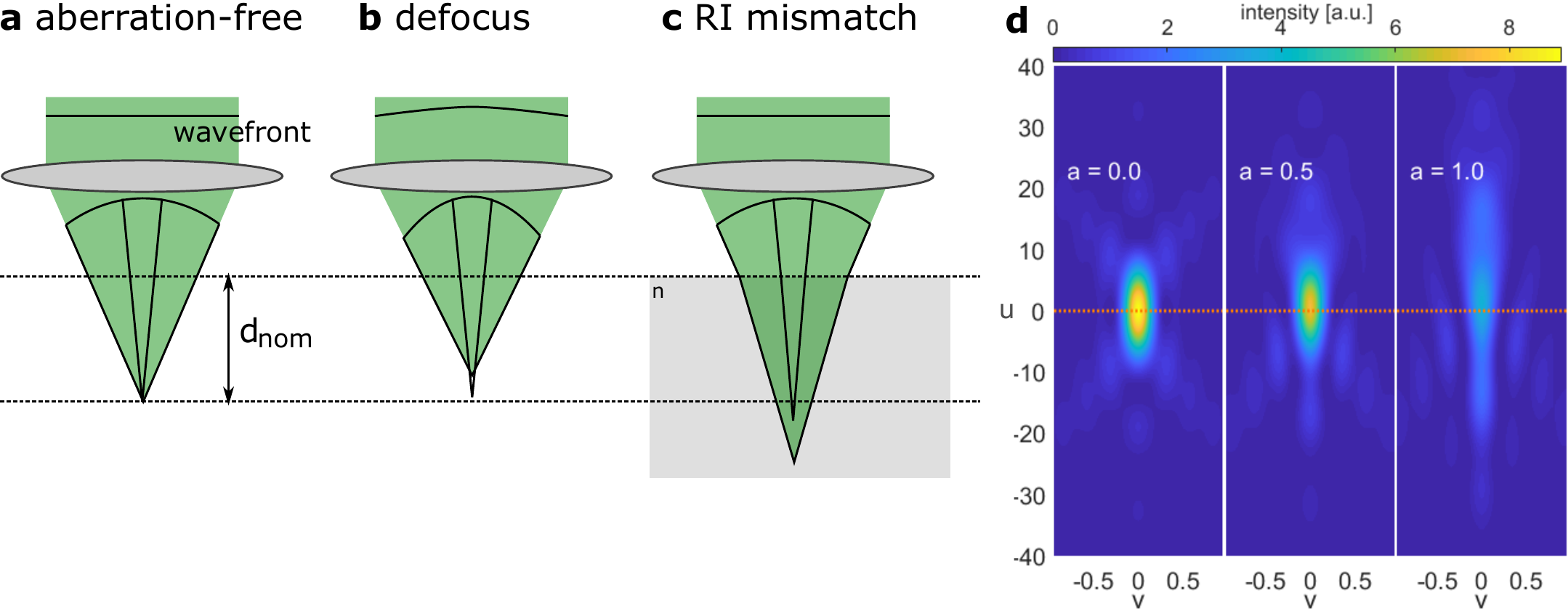}
	\caption{\label{fig:motivation} Illustration of a diffraction-limited focus (\textbf{a}) and sources for spherical aberration (\textbf{b}-\textbf{c}) in adaptive microscopy and its effect on the illumination PSF of a confocal microscope (\textbf{d}). \textbf{a} Optimized lens yields a diffraction-limited focal spot for a propagating plane wavefront. \textbf{b} Aberrations induced by defocused optical system, when operating the adaptive lens outside design configuration. \textbf{c} Refraction index~(RI) mismatch causes spherical aberration in the focal spot. \textbf{d} Diffraction-limited focal spot, shown in normalized optical coordinates $u$~(axial) and $v$~(radial), sectional planes through focal spot. The orange line denotes the nominal focal plane.
	}
\end{figure}

The high tuning range and inertia-free axial scanning capability of adaptive lenses led to their application for axial scanning in several microscopic techniques such as confocal microscopy\cite{Koukourakis:14,Szulzycki2018,Duocastella14,Konstantinou2016}, two-photon microscopy\cite{Grewe2011, Piazza2017}, structured illumination microscopy\cite{Philipp2016hilo, Hinsdale:15, Lin2016,Chen2017}, SPIM\cite{Fahrbach2013,Duocastella2017} as well as standard widefield microscopy\cite{Martinez-Corral2015, Nakai2015, Shain2017}. However, these adaptive optical systems can only be optimized for one specific focal length of the adaptive lens. Thus, a diffraction-limited focal spot as illustrated in~\cref{fig:motivation}a is only achieved at the design focal length of the adaptive lens. Tuning the adaptive lens introduces spherical aberrations that lead to a broadening of the focal spot as depicted in \cref{fig:motivation}b. In microscopic applications, this focal spot broadening corresponds to a rapid decrease of the lateral and in particular axial resolution when operating the adaptive lens outside the design focal length\cite{Fuh2015, Koukourakis:14}. As a result, the tuning range of adaptive lenses cannot be completely used for applications requiring high optical quality. Thus, the compensation of these aberrations is necessary to fully exploit the potential of adaptive lenses. 

Another limitation of spatial resolution in microscopy are spherical aberrations that are induced by specimens itself due to refractive index mismatch (see \cref{fig:motivation}\textbf{c}), leading to a deteriorating imaging quality with increasing penetration depth into a specimen. Multi-actuator adaptive lenses\cite{Bueno2018}, spatial light modulators and deformable mirrors have been applied for aberration correction\cite{booth2014adaptive,Ji2017,Bueno2018,Radner2015,Ji2010a} and scattering compensation\cite{Gigan2017b,Vellekoop2007, Mosk2012, Kuschmierz2018}. While these adaptive optics-based techniques enabled imaging of biological specimens at remarkable resolution and depth, they all come with their specific challenges: Using spatial light modulators in reflection mode~\cite{Ji2010a} or deformable mirrors~\cite{booth2014adaptive} requires beam-folding and, thus, leads to bulky optical setups and hampers their assembly to existing optical systems such as commercial microscopes. In contrast, transmissive spatial light modulators and multi-actuator adaptive lenses can be integrated into optical systems in a collinear geometry. However, their high amount of degrees of freedom require complex control or calibration strategies. While spatial light modulators or deformable mirrors that are used for aberration correction or scattering compensation in optical microscopes, can additionally be used for axial scanning, the tuning range of the refractive power of these beamforming devices is usually much lower than the ones of adaptive lenses, or, when implemented as a Fresnel lens comes with losses in optical quality.

At the same time, the high-order aberration correction and scattering compensation is not required for the measurement of semi-transparent specimens with diameters in the low millimeter range, such as zebrafish embryos. Zebrafish embryos and larvae are used as model organism in developmental biology and toxicology~\cite{Bambino2017}. Particularly transgenic models are highly attractive, since they allow the visualisation of specific structures of interest using reporter-gene driven fluorescence. Until around 6-8 days post fertilsation they are almost transparent. Thus, scattering does only play a minor role due to the high mean free path in semi-transparent specimen. Hence, unwanted contributions from scattered photons can be efficiently suppressed by the pinhole of a confocal microscope\cite{Gu1996} up to penetration depths of several hundreds of microns into biological tissues\cite{Badon2017}. Since the diameter of zebrafish embryos is typically as high as \SI{1}{mm}, spherical aberrations are the key limitation for spatial resolution. In order to achieve sufficient imaging quality, thick biological specimen are usually elaborately prepared for optical microscopy, including slicing~\cite{Opitz2012} or clearing~\cite{Richardson2015}. 

To summarize, the main limitation of the spatial resolution in optical microscopy deep into semi-transparent biological specimen employing adaptive lenses, are spherical aberrations either induced by the specimen or by the optical system due to the adaptive lens itself. Contrary to existing solution with an enormous amount of degrees of freedom, we propose a minimalistic solution with two degrees of freedom -- axial displacement and induced spherical aberrations -- to achieve the same goal of diffraction-limited axial scanning deep inside biological specimens. 
To demonstrate the potential of our approach, we use the adaptive lens in a confocal microscope. The optical sectioning capability of the confocal microscope suppresses efficiently out of focus scattering, leaving spherical aberration as the main contributor to specimen-induced aberration. Using the bi-actor adaptive lens, we attempt to realize axial scanning and reduction of spherical aberration at the same time for each focus position, leading to diffraction-limited foci deep into biological specimen.

\section{RESULTS}

\subsection{Adaptive confocal microscope with spherical aberration correction}
\begin{figure}
	\centering
	\includegraphics[width=.7\textwidth]{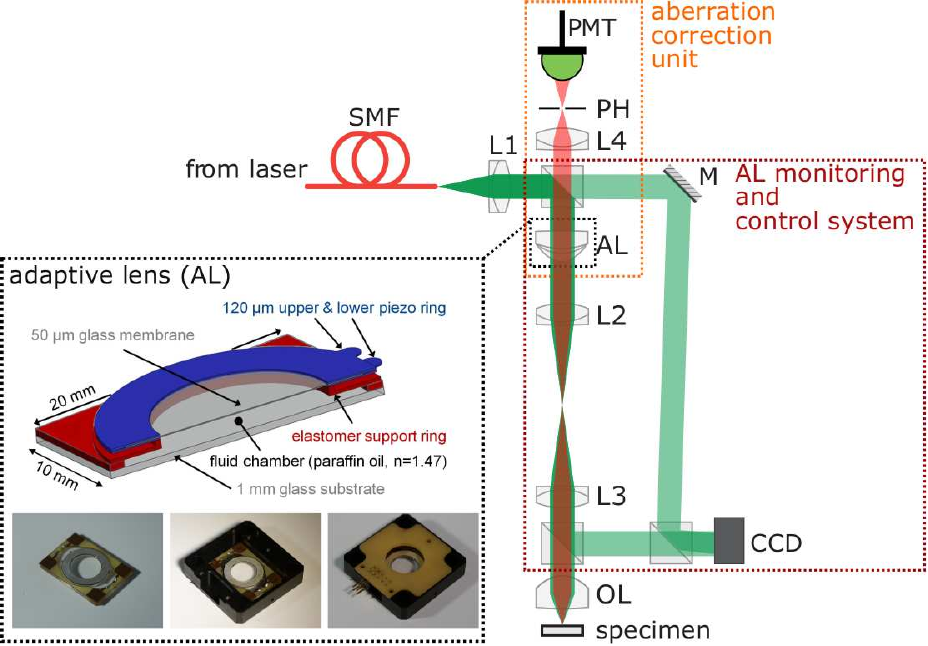}
	\caption{\label{fig:CM} Illustration of confocal microscope with axial scanning and aberration correction unit.  For reasons of clarity only shown with fluorescence, reflection geometry is analogous. \textbf{Inset:} Adaptive lens principle and photographs of the adaptive lens with varying amounts of housing. SMF, single mode fiber; L, lens; OL, objective lens; AL, adaptive lens; M, mirror; PH, pinhole; PMT, detector; CCD, camera. }
\end{figure}

The setup of our custom-build adaptive confocal microscope is illustrated in \cref{fig:CM}.
As the adaptive lens serves both for axial scanning and for spherical aberration compensation, the placement of the adaptive lens in the optical setup is crucial: To ensure axial scanning without inducing additional (defocus) aberration, the adaptive lens has to be placed in telecentric configuration with the objective lens~(OL) of the microscope. 
Additionally, the adaptive lens has to be in a conjugate plane to the focal plane of lens L4 before the detection pinhole, to enable \emph{direct} aberration sensing and compensation. The microscope is extended by an additional Mach-Zehnder-interferometer for phase measurements, whereby an off-axis configuration is used. A Zernike decomposition of the reconstructed phase is conducted and the coefficients of the defocus and spherical aberration terms, $Z_\text{defocus} = \sqrt{3}\left(2\rho^2-1\right)$ and $Z_\text{SA}=\sqrt{5}\left(6\rho^4-6\rho^2+1\right)$, respectively, are used to control the adaptive lens.

Our adaptive lens is an advanced version of the lens introduced in\cite{wapler2014compact} and consists of a \SI{50}{\micro m} thin glass membrane, which is glued in-between two piezo rings (Fig.~\ref{fig:CM}, inset). The glass membrane ensures robustness towards gravity effects that occur when using more flexible membrane materials, such as PDMS. The piezo-glass sandwich seals a fluid chamber that is filled with transparent paraffin oil. The piezo rings enable a direct deformation of the glass membrane. In this way we can not only control the focal power of the lens but also its spherical behavior.  To compensate creeping and hysteresis effects of the piezoelectric actuators, we implement a control system based on wavefront measurements. The wavefront is determined using digital holography in an off-axis Mach-Zehnder-configuration. Note that the wavefront after the relay system (L2 and L3) is measured and used for the closed loop control to also compensate defocus and spherical aberrations introduced by the defocus optical design. 

\subsection{Spherical aberration sensing and correction employing adaptive lens}\label{sec:SAcorrection}

We adapt the protocol first introduced by M.~Booth et al.\cite{Booth2002} for correction of specimen-induced aberrations in a confocal microscope. The aberration correction unit of the adaptive microscope consists of the adaptive lens, the pinhole and the PMT as illustrated in \cref{fig:CM}. For the measurement of the specimen-induced spherical aberration $aZ_\mathrm{SA}$, the adaptive lens is used to deliberately induce the aberrations $(b \pm \Delta b)Z_\mathrm{sp}$, whereby time division multiplexing is applied and $b=b(f_\mathrm{AL})$ is the average supported spherical aberration at a specific focal length $f_\mathrm{AL}$ of the adaptive lens. 

The difference of the signals obtained behind the detection pinhole serves as the response curve to determine the corresponding Zernike coefficient of the wavefront. 

The illumination point spread functions of a confocal microscope with inserted bias aberrations $\pm bZ_\mathrm{SA}(r, \theta)$ equals 
\begin{align}
H_\pm(v, \phi) &= \left|\F\left\{\exp\left(iaZ_\mathrm{SA}(r, \theta) \pm i b Z_\mathrm{SA}(r, \theta) + i \frac{ur^2}{2} \right)\right\}\right|^2 
\end{align}
with the specimen-induced aberration $aZ_\mathrm{SA}(r, \theta)$ and the normalized axial and radial coordinates $u$ and $v$, respectively, defined by $u = \frac{8\pi n}{\lambda} \sin^2(\alpha/2) \cdot z$ and $v = \frac{2\pi}{\lambda} n \sin(\alpha) \cdot r$. Here, $\alpha$ is the angle to the optical axis, $r$ and $z$ are the radial and axial coordinates in real space. 

When considering a fluorescent sheet object lying perpendicular to the optical axis as a specimen and assuming equal excitation and emission wavelengths, the detected signal after the pinhole equals
\begin{align}
W_\pm = \int_0^{2\pi} \int_0^{v_p} (H_\pm * H_\pm) v \dd{v} \dd{\phi}
\end{align} 
whereby $*$ denotes the correlation operator and $v_p$ is the radius of the pinhole. 

The output signal of the modal wavefront sensor is obtained by the difference between the signals with negatively and positively aberrated wavefront $\Delta W = W_+ - W_-$. The measurement (and correction) of the Zernike coefficient $a$ induced by the specimen is realized iteratively using
\begin{align}
c_{n+1} = c_n  + \gamma\Delta W(c_n+b, c_n-b), \label{eq:iterative}
\end{align}
with the initial aberration correction coefficient $c_0=0$ and the gain $\gamma$ that is selected by the user depending on the desired convergence rate and Zernike coefficient range. 

\subsection{Adaptive lens characteristics}
\begin{figure}
	\centering
	\includegraphics[width=0.95\textwidth]{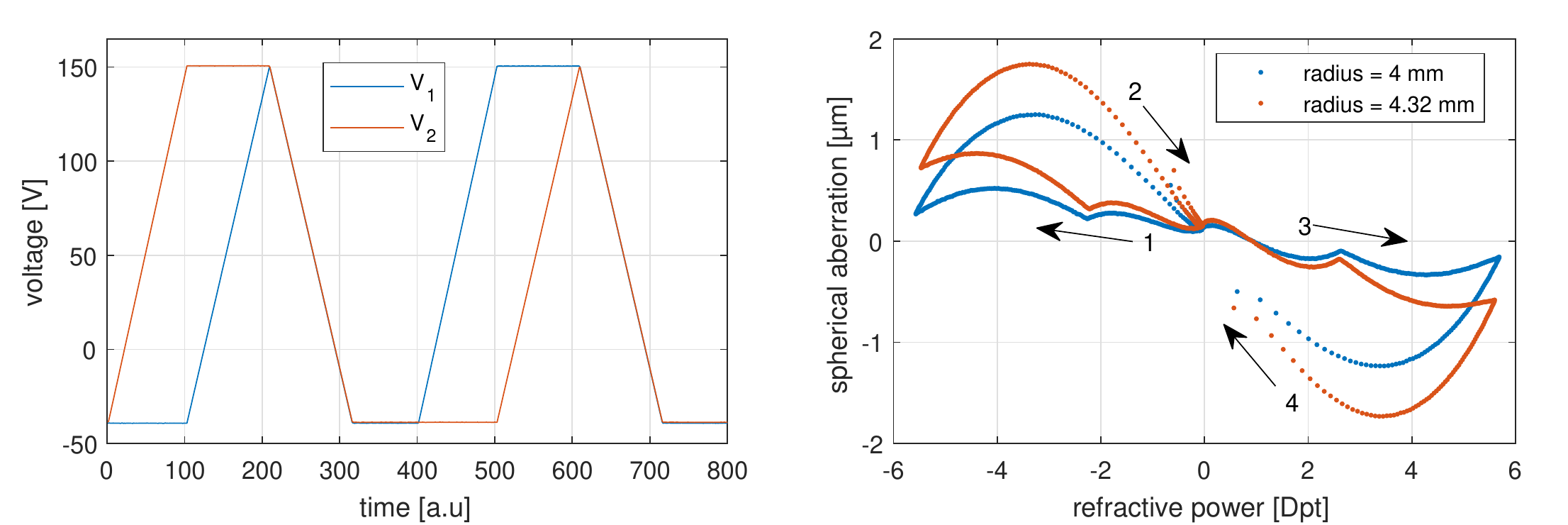}
	\caption{\textbf{a} Typical voltage trajectories for the actuation of the adaptive lens that illustrate the full possible operation region of refractive power and spherical aberration tunability. \textbf{b} Spherical aberrations induced by the adaptive lens in relation to its refractive power, evaluated for a used aperture of radii \SI{4}{mm} and \SI{4.32}{mm}. The arrows and corresponding numbers illustrate the temporal course.}
	\label{fig:ALresponse}
\end{figure}

\cref{fig:ALresponse}b shows an example of a response of an adaptive lens, when actuated with voltages that outline the supported operation region of the lens (\cref{fig:ALresponse}a). The adaptive lens shows a pre-deflection due to an internal strain caused by the manufacturing process and a meta stable jump occurs, where the lens flips from the counter side to the preferred stable pre-deflected side.
The characteristic of the spherical-aberration vs. defocus curve suggests, that the lens is able to compensate spherical aberrations induced by refractive-index mismatch, but does not support zero induced aberrations over the full focal power range. As a result, the optical design of the microscope has to be adapted to allow for the full exploitation of the scanning range of the adaptive lens. Our approach to address this task is to induce a positive spherical bias aberration in the optical system by choosing a front lens with design wavelength of \SI{488}{nm}, when actually using a wavelength of \SI{532}{nm}. To shift the spherical aberrations for negative refractive powers by a negative offset, a coverglas is only used for positive refractive powers of the lens. The slope of the spherical aberration versus defocus characteristic can be steepened by illuminating a larger area of the adaptive lens, compare \cref{fig:ALresponse}b.

\subsection{Diffraction-limited axial scanning in free space}
To demonstrate diffraction-limited axial scanning in free space, we use a mirror as specimen. We measure the axial response of the confocal microscope, i.e. the intensity measured behind the pinhole as a function of the axial mirror position $z$. The full width half maximum (FWHM) of the axial response serves as a figure of merit for the axial resolution.

In the diffracted-limited case, the theoretical axial resolution\cite{Amos2012a} of a confocal microscope under the assumption of an infinitely thin specimen is given by
\begin{align}
\text{FWHM}\left(P_\mathrm{AL}\right) = \frac{0.67\lambda}{n-\sqrt{n^2-\mathrm{NA}^2\left(P_\mathrm{AL}\right)}}\label{eq:FWHM} 
\end{align}
with the numerical aperture NA and the index of refraction $n$ of the surrounding medium.  

\begin{figure}
	\centering
	\includegraphics[width=.57\linewidth]{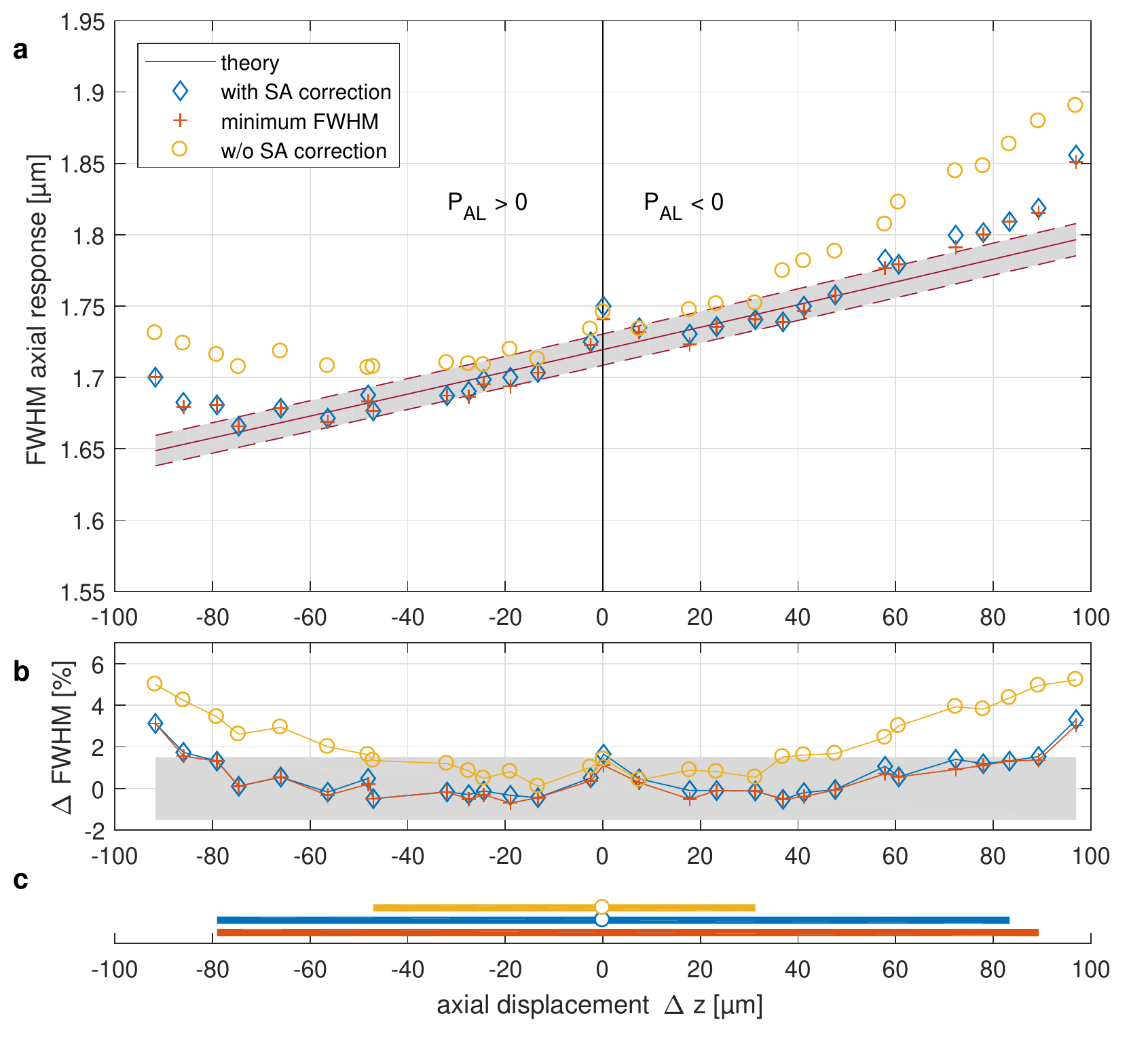}
	\caption{\label{fig:freeSpaceScanning} \textbf{a} FWHM of axial responses of a scanned focus in air with and without spherical aberration correction as a function of the axial displacement of the focal position. A positive refractive power $P_\mathrm{AL}$ corresponds to a negative focal displacement $\Delta z$ and vice versa. As a reference, the theoretical and the minimum FWHM obtained at a given focal length of the adaptive lens are shown. \textbf{b} Relative deviation from measured FWHM to diffraction-limit. The gray area denotes \SI{\pm 1.5}{\%} thresholds. \textbf{c} Regions in which diffraction-limited axial scanning is achieved. White circles denote points where diffraction-limit is not achieved.
	} 
\end{figure}

Hence, the axial resolution of the adaptive confocal microscope only depends on the effective numerical aperture according \cref{eq:FWHM}. The effective numerical aperture is the combined effect of the adaptive lens~(AL) and the objective lens~(OL), that are mapped on conjugate planes by a 4f lens system, see \cref{fig:CM}. Using thin lens approximation, the effective refractive power that defines the effective numerical aperture is $P_\mathrm{eff}=P_\mathrm{AL}+P_\mathrm{OL}$. Increasing the effective numerical aperture yields an axial displacement $\Delta z$ of the focal spot of the confocal microscope in  opposing direction. To account for misalignment of the 4f-system and initial refractive power of the adaptive lens, we determine the effective numerical aperture by measuring the distance between focal spot and the surface of the objective lens using the motorized stage. As a consequence, the theoretical FWHM increases almost linearly with the axial displacement $\Delta z$ of the focal spot, having values between \SI{1.65}{\micro m} and \SI{1.80}{\micro m} over an axial scanning range of \SI{189}{\micro m}, as shown in \cref{fig:freeSpaceScanning}a. The error margin is calculated on basis of the measurement uncertainty of the effective numerical aperture without actuating the adaptive lens and yields about \SI{22}{nm} (full width) over the measured axial range, see \cref{fig:freeSpaceScanning}a.

For an axial scanning system containing a conventional adaptive lens without aspherical tunability, we expect a worsened axial resolution compared to the diffraction-limited case due to operating the objective lens (OL) outside its design operation mode (e.g. illumination with a converging or diverging instead of a plane wave) and often spherical aberrations induced by the adaptive lens itself. We actuate our adaptive lens with approximately equal actuation voltages for both pizeoactors, to mimic a conventional adaptive lens (i.e. $V_1=V_2$). As expected, the axial resolution degrades by up to \SI{5}{\%}, with the deviation being largest for  high absolute refractive powers of the adaptive lens. As a result, only \SI{78}{\micro m} out of the total axial scanning range of \SI{189}{\micro m} are diffraction-limited when considering all FWHMs with less than \SI{1.5}{\%} relative deviation as diffraction-limited, see \cref{fig:freeSpaceScanning}c. 
	
If we actuate the adaptive lens with spherical aberration compensation, the axial resolution improves as illustrated in \cref{fig:freeSpaceScanning}. Here, the focal spot is axially diffraction-limited over an axial tuning range of \SI{162}{\micro m} is diffraction-limited, as shown in \cref{fig:freeSpaceScanning}\textbf{c}. Note that the diffraction-limit is not achieved for zero refractive power of the adaptive lens due to the low tuning range of the spherical aberration at this point, see \cref{fig:ALresponse}. In comparison with the uncorrected case, the spherical aberration correction yields an increase of the diffraction-limited scanning range by a factor of more than two. 

As a measure of the quality of the spherical aberration correction algorithm, we also measure the minimal FWHM at each fixed focal length, see \cref{fig:freeSpaceScanning}. In 12 out of the 29 measured axial responses, the axial resolution with spherical aberration correction is exactly the minimum FWHM. The mean difference between the FWHMs obtained with spherical aberration correction and the minimal FWHMs when tuning the spherical aberration at constant defocus amounts to \SI{2.6}{nm}. The resulting diffraction-limited region is \SI{168}{\micro m}. The marginal deviation between the FWHM with spherical aberration correction and the minimal FWHM might be a result of not fully compensated piezo creeping during the measurement of the axial resolution. In conclusion, the spherical aberration correction exploits the potential of the adaptive lenses to a large extent.

\subsection{Specimen-induced aberration correction at a phantom specimen}

\begin{figure}
	\centering
	\includegraphics[width=.7\linewidth]{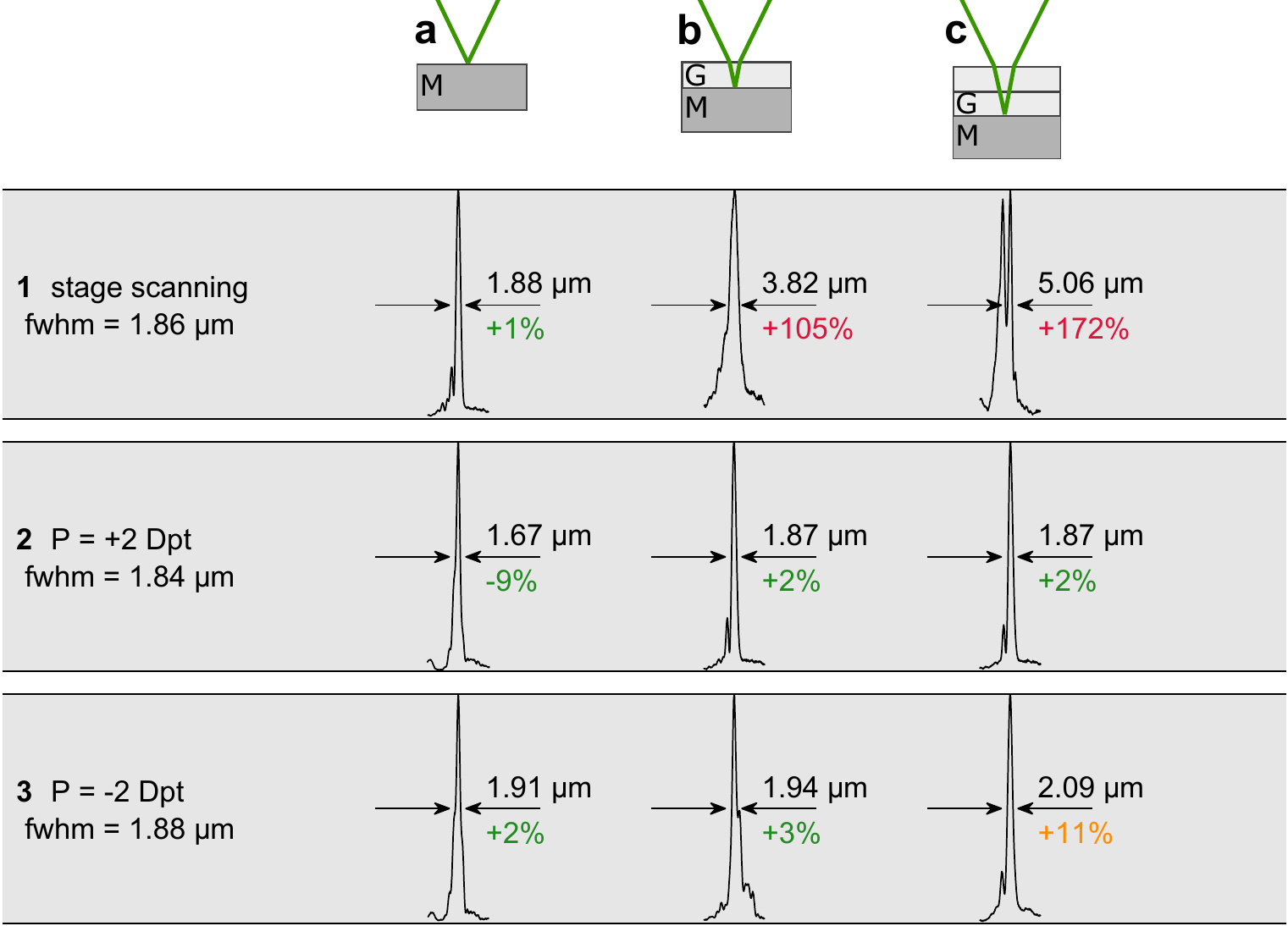}
	\caption{\label{fig:ascan_model} \textbf{a}-\textbf{c} Illustration of phantom specimen. Corresponding normalized axial response curves employing stage scanning without spherical aberration compensation (top row), with aberration correction at refractive power \SI{2}{Dpt} (middle) and \SI{-2}{Dpt} (bottom). The theoretically calculated FWHMs for the diffraction-limited cases are denoted in the left row. The measured FWHM are stated at the corresponding axial response curves. The deviation of the measured FWHM to the diffraction-limit are expressed in percentages.  M, mirror; G cover glass. } 
\end{figure}

In order to test the spherical aberration correction capability of our adaptive microscope for specimen-induced aberrations, we use three types of phantom specimen to mimic index of refraction mismatch. The phantom specimen consists of a mirror and up to two coverglasses (see header row in \cref{fig:ascan_model}) with a thickness of \SI{170}{\micro m} and a refractive index of $n_\text{BK7}=1.52$ at $\lambda = \SI{532}{nm}$ (BK7, Schott AG). Consequently, measurements at depths $d_\text{BK7}$ of \SI{170}{\micro m} and \SI{340}{\micro m} in the phantom specimen are conducted. As the index of refraction of zebrafish embryos is about $n_\text{embryo}=1.35$ according~\cite{Sampath2010} and thus significantly deviates from the the one of BK7, we use the optical path length $\text{OPL}=nd$ to transfer the observations of this section to zebrafish embryos. As the optical path length is invariant of the surrounding medium, the to BK7 corresponding depths inside the zebrafish embryos are $\frac{n_\mathrm{BK7}}{n_\mathrm{embryo}}d_\mathrm{BK7}$, which yields propagation depths of \SI{189}{\micro m} and \SI{378}{\micro m} for one and two coverglasses, respectively. As the thyroid of zebrafish embryos is not expected to be position much more than \SI{200}{\micro m} beneath the surface~\cite{Opitz2012}, it is our goal to demonstrate diffraction-limited axial scanning up to this depth.

As a reference, we record the axial response of the confocal microscope with our adaptive lens being in its ground state at $V_1=V_2=0$ for different positions of the mirror, see first row in \cref{fig:ascan_model}. When using a mirror without coverglasses as specimen, the axial resolution is limited by the numerical aperture $\mathrm{NA}=0.6$ of the objective lens according \cref{eq:FWHM}, which yields $\text{FWHM} = \SI{1.86}{\micro m}$. The FWHM of the measured axial response equals \SI{1.88}{\micro m} which deviates only by \SI{1}{\%} from the theoretically obtained diffraction limit, see \cref{fig:ascan_model}, cell \textbf{1a}. As expected, the FWHM of the axial response increases with increasing propagation depth by a factor of up to $2.7$ (or an increase of \SI{172}{\%}) due to the refractive index mismatch. 

We demonstrate the spherical aberration correction capability inside a specimen exemplary at refractive powers $P_\mathrm{AL} = \pm \SI{2}{Dpt}$ to ensure that the procedure works both for wavefronts that are diverging as well as converging when reaching the objective lens. The theoretical diffraction limits according to \cref{eq:FWHM} change to \SI{1.84}{\micro m} and \SI{1.88}{\micro m} for $P_\mathrm{AL} = \pm \SI{2}{Dpt}$, respectively, due to the influence of the refractive power of the adaptive lens on the effective numerical aperture of the optical system. The FWHMs obtained from the measured axial responses lay in the expected region, as they are \SI{9}{\%} below or \SI{2}{\%} above the theoretically obtained diffraction-limit for $P_\mathrm{AL} = \pm \SI{2}{Dpt}$, respectively, as shown in \cref{fig:ascan_model}~(cells \textbf{2a} and \textbf{3a}). When using the phantom specimen consisting of one coverglass on top of the mirror, we expect no change of the FWHM as the effective numerical aperture remains constant. The FWHM of the measured curves fulfill this expectation as they are only \SI{2}{\%} to \SI{3}{\%} above the theoretically obtained diffraction limit, as apparent from \cref{fig:ascan_model}~(cells \textbf{2b} and \textbf{3b}). In the case $P_\mathrm{AL} = +\SI{2}{Dpt}$, this trend continues for the phantom specimen with two coverglasses on top of the mirror and the measured FWHM is again only \SI{2}{\%} above the diffraction-limit. In the case of $P_\mathrm{AL} = -\SI{2}{Dpt}$, however, the FWHM obtained from the measured axial response is \SI{11}{\%} higher than the diffraction-limit, i.e. the focal spot is only near-diffraction limited in this specific configuration, see cell \textbf{3c} in \cref{fig:ascan_model}. 
As apparent from rows \textbf{b} and \textbf{c} in \cref{fig:ascan_model}, the axial broadening of the focal spots \SI{170}{\micro m} deep into the phantom specimen has been reduced from \SI{105}{\%} by a factor of 50 and 35 to a broadening of \SI{2}{\%} and \SI{3}{\%} for refractive powers $P_\mathrm{AL} = \pm \SI{2}{Dpt}$, respectively, compared to the reference measurements with the stage scanning system. At a propagation depth of \SI{340}{\micro m} into the specimen, the axial broadening was reduced from \SI{172}{\%} by a factor of 86 and 17 to \SI{2}{\%} and \SI{11}{\%} for refractive powers $P_\mathrm{AL} = \pm \SI{2}{Dpt}$, respectively, compared to the reference measurements with the stage scanning system.

\subsection{Correction of specimen-induced aberrations at in-vivo measurements of zebrafish embryo thyroids}
\label{sec:specimenInduced}

After the demonstration of diffraction-limited axial scanning inside phantom specimen, we now use our adaptive confocal microscope to measure reporter gene-driven fluorescence in the thyroid gland of a zebrafish embryo with and without spherical aberration correction. As was shown in the last section,  diffraction-limited axial scanning is possible for phantom measurements up to at least \SI{340}{\micro m} deep in BK7. This corresponds to a penetration depth of \SI{378}{\micro m} in zebrafish embryos due to the invariance of the optical path length from the surrounding medium. As the diameter of zebrafish embryos typically does not exceed \SI{1}{mm} and the thyroid is not located more than \SI{200}{\micro m} beneath the surface~\cite{Opitz2012}, we expect to achieve  diffraction-limited or near-diffraction-limited focal spots by spherical aberration correction for measurements of the reporter gene-driven fluorescence in the thyroid gland of zebrafish embryos.

\begin{figure}
	\centering
	\includegraphics[width=\textwidth]{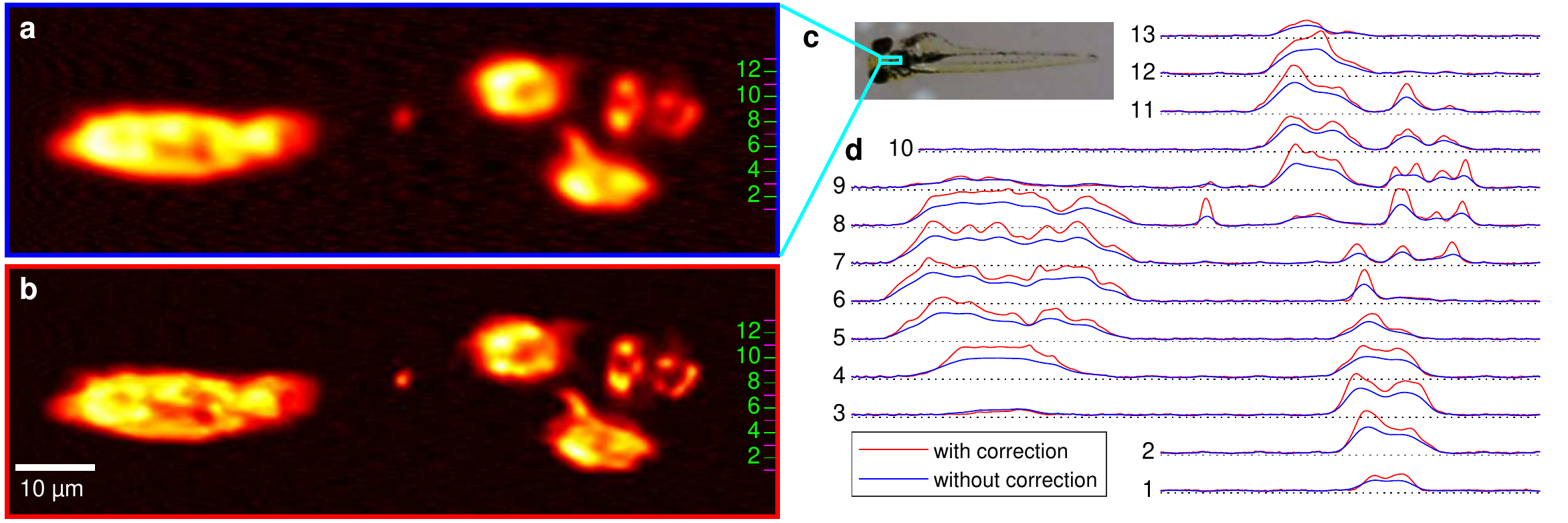}
	\caption{\textbf{a}+\textbf{b} Images of a zebrafish embryo at 110 hours post fertilization with reporter gene-driven fluorescence in the thyroid gland using the adaptive microscope without (\textbf{a}) and with (\textbf{b}) spherical aberration correction. The images were normalized separately to show the intensities using the full color range. \textbf{c} Measured region is approximately marked in standard widefield image of the whole embryo. \textbf{d} Line profiles of the gray values of the non-normalized versions of (\textbf{a}+\textbf{b}) are taken at specific rows, where the positions of lineouts are indicated by the green and black numbers in (\textbf{a}+\textbf{b}) and (\textbf{d}), respectively. Lineouts of fluorescence images indicate that aberration correction leads to increased signal intensity and better optical sectioning. Ventral views, anterior is on the left (\textbf{a}-\textbf{d}). Dimensions are $(100\times308)\SI{}{\micro m^2}$ (\textbf{a}-\textbf{b}), about $(1.5\times5)\SI{}{mm^2}$ (\textbf{c}) and $(70\times308)\SI{}{\micro m^2}$ (\textbf{d}).}
	\label{fig:zebrafish}
\end{figure}

In order to investigate the effect of our spherical aberration correction procedure on the measurement at zebrafish embryos, we acquire a fluorescence image of the thyroid without spherical aberration correction as a reference. Please keep in mind, that we operate our microscope to induce a slight spherical aberration bias to already have a rough compensation of the index of refraction mismatch. Whitout that pre-compensation of spherical aberrations, we had difficulties acquiring a reference image as the thyroid dimensions are very tiny compared to the overall zebrafish dimensions, see \cref{fig:zebrafish}\textbf{c}. As a result, the reference image as shown in \cref{fig:zebrafish}a is already pre-compensated for spherical aberration. To ensure that photobleaching and phototoxity do not compromise our results, the uncorrected the images without applying the spherical aberration correction procedure are acquired before the ones with correction. 
The follicles of the thyroid are already perceivable distinctly in the pre-compensated reference image, while the substructure cannot be resolved detailedly yet. In contrast, a substructure of the follicles is apparent in the image acquired with correction of specimen-induced aberrations, which allows to get a better understanding of the spatial distribution and extension of the thyroid follicles (\cref{fig:zebrafish}\textbf{b}). The images were normalized separately to show the intensities using the full color range. 

The correction of specimen-induced aberrations also manifests itself in an increased contrast and overall signal strength as directly apparent from comparing line profiles of the image with specimen-induced aberrations and the reference as shown in \cref{fig:zebrafish}\textbf{d}. The maximum intensity is increased by a factor of about two. The contrast defined as the quotient of standard deviation and mean value of the intensity values, distinctly evaluated over the follicles yields contrast increases by factors between 1.5 and 2.8, depending on the position and size of the evaluation window. The line profiles also highlight the arise of follicle substructures when applying the correction of spherical aberrations, see e.g. left side of line profiles number 7, 8 and 9.

\section{DISCUSSION}

The main result of this paper is that our bi-actuator lens is capable to axially scan a focus while maintaining an axially diffraction-limited spot both in free space and deep inside specimens, even when employed in a multi-component imaging system such as a confocal microscope. This is achieved by the correction of spherical aberrations that are either induced by the specimen or by the optical system due to the axial scanning procedure.  

We like to highlight, that this goal of axially diffraction-limited scanning is achieved by only two degrees-of-freedom which constitutes the minimal possible solution. This is in sharp contrast to the typically more than a million degrees-of-freedom present in spatial light modulators and deformable mirrors that are currently the gold standard in adaptive-optics-based aberration correction procedures. Further, inserting only the adaptive lens (and two relay lenses) into an existing confocal microscope enables the measurement and correction of spherical aberrations yielding a compact optical setup without additional beamfolding and the need for complex adjustment of several additional optical components. While we added an additional Mach-Zehnder-interferometer to our microscope to implement a control system for compensating the hysteresis of the piezoactors of the adaptive lens, a direct control of the piezoactors by measuring the charge or capacity is also possible and would not require any additional optics.

The spherical aberration versus defocus characteristics of the adaptive lens (\cref{fig:ALresponse}b) has a positive effect when focusing into a specimen as the default lens surface at a certain defocus (i.e. equal actuation voltages) contains an inherent spherical aberration term that is already opposed to specimen-induced aberrations. As a result, the acquired images of zebrafish embryos with reporter gene-driven fluorescence in the thyroid gland have already a decent quality (\cref{fig:zebrafish}a), even without explicit application of the spherical aberration correction algorithm. Applying the correction algorithm yields further increased contrast and by a factor of approximately two increased fluorescence signal (\cref{fig:zebrafish}\textbf{b}+\textbf{d}). A limitation of our microscope is, however, that it cannot adequately compensate non-symmetric aberrations due to e.g. non-perpendicular illumination of the zebrafish embryos or embryos with particularly uneven surfaces. As a result, not every imaging attempt of the thyroid of zebrafish embryo was successful, in agreement with former observations\cite{Turcotte2017}.

To summarize, the proposed bi-actuator adaptive lens enables a high-usability, compact and potentially low-cost method to extend an existing imaging system with diffraction-limited axial scanning capability. We hope, the presented approach leverages success for inertia-free, diffraction-limited axial scanning employing adaptive lenses with spherical aberration compensation for applications with high demands regarding optical quality and penetration depth. The presented approach is transferable to confocal techniques such as confocal Brillouin spectroscopy\cite{Edrei2018}, two/multi-photon microscopy with spherical aberration compensation~\cite{Matsumoto2015} and high-resolution laser-machining.

Future work will focus on optimizing the adaptive lens characteristic (\cref{fig:ALresponse}b) to allow zero spherical aberration over the full defocus range. In contrast to the current situation, this will enable diffraction-limited axial scanning in free space at both negative and positive refractive powers of the adaptive lens without the need to adapt the optical system accordingly.

\section{MATERIALS AND METHODS}

\subsection{Characterization of the adaptive lens surface}

The adaptive lens was characterized by scanning the lens surface with a confocal distance sensor in combination with a xy-stage while the lens was actuated using a function generator. The reconstructed time-dependent three-dimensional membrane shape was then used to obtain the defocus and spherical aberration Zernike coefficients of the resulting optical path difference (shown in \cref{fig:ALresponse}) under consideration of the refractive index of the adaptive lens fluid $n_\text{AL}$. The refractive power of the adaptive lens was calculated from the defocus coefficient analogously to \cref{eq:refractivePower}.

\subsection{Phase measurements and Zernike decomposition} \label{sec:MethodsZernike}

Phase measurements were conducted with an off-axis Mach-Zehnder interferometer, whereby the hologram was recorded by CCD camera (pco.pixelfly usb, 1392 x 1040 pixel, pixel size \SI{6.5}{\micro m}, 14 bit dynamic range). The angular spectrum beam propagation method\cite{Verrier2011} is used to determine the phase of the wavefront at the conjugate plane of the adaptive lens. A Zernike decomposition of the phase is conducted in order to extract defocus and spherical aberration coefficients. 
The refractive power $P$ of the adaptive lens is obtained from defocus term of measured wavefront using 
\begin{align}
P =  \left(n_\mathrm{AL}-1\right)\frac{4\sqrt{3}\lambda \alpha_2^0}{a^2} \label{eq:refractivePower}
\end{align}
with the refraction index $n_\mathrm{AL}=1.48$ for the surrounding medium and the lens fluid, respectively. $a$ is the radius of the aperture used for Zernike evaluation and $\alpha_2^0$ is the defocus Zernike coefficient. A derivation of \cref{eq:refractivePower} can be found in \cite{philippa2017spherical}. 
It should be noted, that the combined effect of the adaptive lens and the two achromatic lenses of the relay system is measured (compare \cref{fig:CM}) and, thus, the control occurs regarding the effective spherical aberration and defocus of those three lenses. Consequently, aberrations due to imperfect optical alignment along the optical axis are partly compensated as well.

\subsection{Control of spherical aberration and defocus induced by adaptive lens}
The creeping and hysteresis effects of the piezoelectric actuators make it necessary to control the actuation process. We implement a control system based on wavefront measurements that also compensates influences from environmental conditions such as temperature and pressure. The Zernike coefficients are determined as described in the previous paragraph. Initial actuation voltages are determined based on a previously recorded look-up-table between actuation voltages and defocus as well as spherical aberration Zernike coefficients. The actuation voltages are then applied to the adaptive lens and the Zernike coefficients of the wavefront are obtained again. The actuation voltages are then refined iteratively based on the gradient of the look-up-table. 

\subsection{Zebrafish embryo preparation}
Zebrafish of the strain tg:mCherry were originally provided by the University of Brussels\cite{Opitz2012}. Embryos were obtained as described by Fetter et al.~\cite{Fetter2015}. For imaging of the thyroid gland embryos at 5 dpf were used. Embryos were anestetised with a tricaine solution (150mg/L, TRIS 26mM, pH 7.5) and embedded in \SI{3}{\%} methyl cellulose to stabilise a dorsoventral position.

\subsection{Fluorescence image acquisition of zebrafish embryos}

The images were obtained by lateral scanning using a Newport XPS system. The $y$-axis is used as the slow, and $x$ is used as the fast axis. The position along the $y$-axis was moved stepwise with steps of \SI{500}{nm}. The velocity of the $x$ axis was set to \SI{2}{mm/s} with an acceleration of \SI{20}{mm/s^2}. The acquisition rate of the Newport xps was set to \SI{4}{kHz}, yielding also a stepwidth of about \SI{500}{nm} after the acceleration process is finished. The so obtained scattered data were resampled on a regular grid with a pixel size of $500\times\SI{500}{nm^2}$ using the \emph{ScatteredInterpolant} class of MATLAB(R) using linear interpolation.

\subsection{Spherical aberration correction procedure at measurements of zebrafish embryos}
Since the mCherry dye is sensitive to photo-bleaching, minimizing the exposure time was a key priority when establishing the correction procedure. Consequently, we only measure the laterally integrated intensity for the iterative aberration correction using~\cref{eq:iterative}. We performed frame-wise aberration correction, because intermittent illumination causes less photo-bleaching compared to continuous illumination\cite{Shaner2014}.

\nolinenumbers
\appendix

\section*{ACKNOWLEDGMENTS}
We thank Markus Finkeldey for several helpful discussions and suggestions regarding building a confocal microscope. We also thank David Leuthold, Helmholtz Centre for Environmental Research - UFZ, for production of zebrafish embryos. Prof. Sabine Costagliola and Dr. Robert Opitz (University of Brussels) are acknowledged for providing the tg(tg:mCherry) transgenic zebrafish strain. The financial support of the Deutsche Forschungsgemeinschaft (DFG) for the projects CZ 55/32-1, Wa 1657/6-1 and SCHO 700/6-1 is gratefully acknowledged. 

\section*{AUTHOR CONTRIBUTIONS} 

KP designed and performed the experiments, derived the models and analyzed the data. FL fabricated the adaptive lenses. StS provided guidance for zebrafish embryo analysis, particular positioning of embryos and microscopic analysis. NK shared code for quantitative phase imaging. MW had the initial idea of the adaptive lens principle, NK and JC initiated the concept of aberration correction in adaptive confocal microscopy and all authors optimized the lens specifications for application in the confocal microscope. KP wrote the manuscript with contributions from FL regarding the adaptive lens fabrication. JC and UW are the principal investigators. All authors read and commented on the manuscript.

\section*{CONFLICT OF INTEREST}
The authors declare no conflict of interest.

\section*{DATA AVAILABILITY}
Experimental data are available from the corresponding author on reasonable request.

\section*{REFERENCES}
\hspace{51mm}

\bibliography{report} 
\bibliographystyle{naturemag}

\end{document}